# Informality, Education-Occupation Mismatch, and Wages: Evidence from India

**Shweta Bahl**

Indian Institute of Management Rohtak, Rohtak,

Haryana, India

e-mail: shweta.bahl@iimrohtak.ac.in;  shweta.bahl28@gmail.com.

**Ajay Sharma (Corresponding Author)**

Indian Institute of Management Indore, India

Fellow, Global Labor Organization

e-mail: ajays@iimidr.ac.in; ajaysharma87@gmail.com.

Phone: +91-731-2439622.

**Abstract**

*This article examines the intertwining relationship between informality and education-occupation mismatch (EOM) and the consequent impact on the workers' wages. In particular, we discuss two issues — first, the relative importance of informality and education-occupation mismatch in determining the wages; and second, the relevance of EOM for formal and informal workers. The analysis reveals that although both informality and EOM are significant determinants of wages, the former is more crucial for a developing country like India. Further, we find that EOM is one of the crucial determinants of wages for formal workers, but it is not critical for informal workers. The study highlights the need for considering the bifurcation of formal-informal workers to understand the complete dynamics of EOM especially for developing countries where informality is predominant.*

**Acknowledgment:** We would like to thank Karthikeya Naraparaju, Deepak Sethia, and seminar participants at the annual conference of Indian Society of Labour Economics, 2018 for their helpful comments and suggestions on the earlier draft of this paper. The usual disclaimers apply.

# 1. Introduction

How do labour markets work have always fascinated academicians and policymakers. One of the critical questions in this context relates to the determination of wages from the perspective of workers, firms, as well as institutions such as government, regulators, and so on. There are two complementary strands in the literature that attempt to explain the wage determination process. The first strand argues that it is the structure of labour market guided by institutions (rules, regulations, social structure and historical path dependence, and so on) that influences the wages in the labour market. The seminal work of Lewis (1954) discussing duality in labour market in terms of the formal and informal sector (and extended by Harris and Todaro (1970), Fields (1990) among others) highlights the importance of this channel.[1] The second strand argues that the characteristics of demand and supply side and their interactions play a crucial role in the process of wage determination. This includes seminal work of Becker (1964) and Mincer (1974) on the human capital theory at one side and the competing job competition model (Thurow, 1975) and assignment model (Sattinger, 1993) on the other.

Further, as an extension to the second strand, an extensive literature in the past three decades has evolved using the argument of (mis-)match between demand and supply characteristics, popularly called education-occupation mismatch (EOM), as an important factor in the wage determination process. EOM is defined as an incongruity between the educational attainment of a worker and educational requirements of her job (Duncan and Hoffman, 1981). There can be three possible match status for a worker considering the definition of EOM: overeducated (when attained education is higher than required education), undereducated (when attained education is lower than required education), and adequately educated (when attained and required education is aligned). This article attempts to examine the intertwining relationship between the two strands of wage determination process, notably informality and EOM and their relative impact on the wages of workers.

---

[1] Other institutions/policies that influence the wages include minimum wage legislation, regulation of working hours, unions and collective bargaining, and so on (see, Boeri and Van Ours (2013) for complete discussion).

Though these two strands have evolved independent of each other, there are considerable similarities that can be observed in their arguments. In the dual labour markets, especially in developing countries, a small number of workers are employed in formal sector which pays higher wages than immensely crowded informal sector (Pradhan and Van Soest, 1995). It is argued that due to high wages offered in the formal sector, there is rationing that causes some of the eligible workers to either remain unemployed or find work in the informal sector (Gindling, 1991). Essentially, we can also term these workers as mismatched, or specifically overeducated for the jobs they are employed in. This consequently results in wage gap among workers with similar observable characteristics. This remains one of the pivotal findings of dual sector models and corresponding empirical results. A similar argument is at the core of the EOM framework. Freeman's analysis indicates that if the number of workers on the supply side of a specific job is higher than the demand for them and the prospective workers have higher attained education than the required; then it is likely to result in overeducation and vice-versa (Freeman, 1976).

The critical difference remains that EOM argues for mismatch at occupation category, whereas dual labour market theory argues for mismatch at the sector level that is formal and informal. Another distinctive feature of the dual labour market and EOM comes from the context in which they are explored. Interestingly, dual labour market theory and its implications are primarily confined to developing countries where informality arises on account of some missing institutions (Ruppert Bulmer, 2018). On the other side, EOM literature has mainly gained traction in the developed countries starting with Freeman's work on overeducation in the United States (Freeman, 1976).[2] This is justified since EOM was not a concern for developing countries two decades ago as compared to other issues like unemployment (O'Higgins, 2003), the transition from agriculture to non-agriculture employment (FAO, 1998), and so on. But, over time with significant public investment on education by developing countries (Hien, 2018), the focus has moved from providing employment to the provision of decent and suitable employment (ILO, 2002). Moreover, unlike the trajectory of developed countries, the informality remains a permanent feature of developing countries.

The arguments above suggest that the literature on either strand is quite rich and has similarities too. However, the strands have evolved mainly in isolation. There are not enough studies that combine the framework of informality and EOM to understand wage determination in the labour market. To the best of our knowledge, only one study examines the relationship between informality and EOM. The research by Herrera-Idárraga et al. (2015) conducts this analysis for Colombia. The study defines informality on the basis of the absence of social security benefits in a job and finds that there is a double penalty in the case of informal employment. Informally employed workers suffer not only lower returns to education but also receive a higher penalty for being mismatched (both overeducated and undereducated). The article, however, does not bring out the relative importance of informality and EOM in explaining the wages of the workers. Besides, informality can arise from two aspects – nature of the job (type of employment) and nature of the firm (sector of work). While focusing only on the distinction in the context of employment, the study overlooks a crucial segment of the labour market – the sector of work being formal or informal.

---

[2] For example, a review study by McGuinness et al. (2018) finds that overeducation is studied for 39 countries in the last decade, out of which, as per the categorization of '*The World Factbook*', 72 percent are developed countries.

This article attempts to fill this lacuna by analysing the intertwining relationship between informality and EOM in the context of developing countries using the case of India. Based on the existing literature, some key issues could be raised. First, the relative importance of informality and EOM in determining wages in the labour market of a developing economy. Second, whether (and how much) difference exists in pecuniary returns to EOM across formal and informal workers. The answers to these research questions stimulate not only a theoretical perspective but also policy-making. For example, if EOM does not attract any penalty for informal workers, then policymakers have an incentive to encourage the acceptance of job offers while ignoring the adequacy of a match between education and occupation. However, if the opposite holds, then informality would be a double-edged sword. Apart from inferior working conditions, it will also demonstrate a wage loss due to EOM.

The contribution of this article is twofold. On one side, it contributes to the understanding of wage determination in developing countries where both informality (Nguyen et al., 2013) and EOM (Quinn and Rubb, 2006) mechanisms are active in wage determination. In particular, this analysis is intended for those countries where the level of informality has not gone down, but education levels have improved rapidly. Therefore, they are likely to face implications from both the mechanisms. India, as an example of a developing country, makes a compelling case for the following reasons. First, as per 2018-19 periodic labour force survey conducted by India's National Sample Survey Office (NSSO), informal workers in India comprise 72 percent of the total labour force. To add to this, the formal-informal sector wage gap and the formal-informal employment wage gap is also quite significant (Unni and Sarkar, 2013). Further, 2018-19 periodic labour force survey reveals that the proportion of working age people (15-59 years) with no formal schooling (21 percent) and higher education (26 percent) is quite substantial and hence, it increases the likelihood of being undereducated and overeducated respectively.[3]

Coming to the second contribution, the study presents a more comprehensive picture of informality. Informality can arise in two aspects. On one hand, from the nature of job (type of employment) that is, in terms of job security, work contract, and social security. The other aspect of informality arises from the nature of firm that is, compliance concerning laws and regulation, paying taxes or not, and so on. In the literature, informality in sector and informality in employment has mostly been considered in isolation from each other. However, it has been found that by creating a dual structure in their respective enterprises, formal firms hire unskilled workers on a contract rather than permanent basis (Ramaswamy, 2012), which consequently leads to informalisation of jobs in formal firms. Considering this, the study bridges the literature between informality in sector and informality in employment by examining the intricacies of both.

---

[3] The corresponding proportions are approximately same for the employed population.

The main results are as follows. First, considering the individual impact of informality and EOM on wages, we find that our results are consistent with the existing literature. Both the informal sector and informally employed workers earn lower wages than their formal counterparts. Also, overeducated workers are penalized (eight percent) in the labour market irrespective of the sector of work. Second, the inclusion of informality in employment alters the magnitude of penalty for being overeducated. Third, segregating the data first by formal and informal sector, we find that EOM is not crucial for informal sector workers. They neither get rewarded for being undereducated nor get penalty for being overeducated. On the other hand, formal sector workers while do not get rewarded for being undereducated get a significant penalty for being overeducated. Lastly, coming to the context of informally employed workers, the study reveals that while they receive penalty for being for being overeducated, it is lower than that of formally employed workers.

The remainder of the article is structured as follows. The following section presents the conceptual framework. Section 2 defines informality and EOM. Section 3 discusses the data source and descriptive statistics. Section 4 entails the empirical methodology. Section 5 discusses the results. The final section concludes the study.

**2. Definition: Informality and Education-Occupation Mismatch**

This section describes the definition of informality with respect to employment (nature of job) and sector (nature of firm) and education-occupation mismatch (EOM).

## 2.1. *Informality*

The definition of informality varies across countries due to a lack of consensus on a theoretical basis and the vast array of activities covered. In the Indian context also, different government agencies used to follow varied definitions and nomenclatures to define the informality. Hence, designing a uniform approach to administer the informality and solve the associated problems was nearly impossible. Considering this, the National Commission for Enterprises in the Unorganized Sector (NCEUS) had set up a task-force to review the existing definitions and come up with the harmonised definition of the informal sector and informal economy. As per NCEUS (2008), "*informal sector consists of all unincorporated private enterprises[4] owned by individuals or households engaged in the sale of and production of goods and services operated on a proprietary or partnership basis and with less than ten total workers*" (pg. 2). In other words, a worker is said to be working in the informal sector if she is working with an unincorporated enterprise that employs less than ten workers. Unincorporated enterprises in our case refer to proprietary or partnership firms. Further, a worker is said to be informally employed if she is "*working in the informal sector or households, excluding regular workers with social security benefits provided by the employers and the workers in the formal sector without any employment and social security benefits provided by the employer*" (pg. 2).

---

[4] Enterprises with type of ownership as proprietary or partnership (NSSO, 2012).

Therefore, if a worker is not entitled to any social security benefits, she will be classified as informally employed irrespective of her employment sector.

The definitions are consistent with the international definitions and framework (NCEUS, 2008). However, the international definitions even though essentially capture the same aspect, are less stringent for the informal sector and more stringent for informal employment. For example, the International Labor Organization (ILO) categorizes a worker in the informal sector if she is working with an unregistered enterprise irrespective of firm size (ILO, 1998). Besides, for a worker to be classified as informally employed, she must be working with an unregistered enterprise, and should not be eligible for social security benefits (ILO, 2003). Hence, while NCEUS estimates for the informal sector are conservative, opposite holds for informal employment.

## 2.2. *Education-Occupation Mismatch*

As mentioned earlier (Section 1), EOM is defined as the mismatch or misalignment between attained education of an individual and required education by her occupation. Therefore, to measure EOM, the prerequisite is to have information on attained and required education. While the former can be obtained from individual-level data, to estimate the latter, the literature provides three methods - worker's self-assessment, job analysis, and realized matches (Leuven and Oosterbeek, 2011)[5]. This study uses

---

[5] *Worker's self-assessment* involves asking the respondents about the educational requirements at their respective jobs (Duncan and Hoffman, 1981) or more directly, their respective match status (over, under or adequately educated) (Chevalier, 2003). *Job analysis* involves careful examination of the

realized matches (RM) approach based on its strengths mentioned below. The rest of the discussion is confined to RM only, for sake of brevity[6].

*Realized matches* is a statistical approach that captures mismatch from the perspective of the labour market and hence, involves an interaction of demand- and supply-side factors. RM gauges a worker's match status by comparing her attained education with the statistical threshold such as mean (Verdugo and Verdugo, 1989) and mode (Kiker et al., 1997). This threshold is calculated using the education profile of all the workers working in a particular occupation. Withal, the hitch is that RM captures the selection and screening of an employee rather than the actual level of education required for a job (Verhaest and Omey, 2006). However, this can also be an advantage as it unveils the intrinsic requirements of the job prevailing at present. The added advantage is that it is widely used (Leuven and Oosterbeek, 2011) and hence, aids in establishing comparisons with the existing studies.

## 3. Data and Descriptive Statistics

This section provides the source and description of data to support the context of this article. The section also highlights the variables required for the identification of informality in sector and employment and education-occupation mismatch (EOM).

### *3.1. Data Source*

---

occupations by the professional job analysts to ascertain educational requirements associated with a particular job (Rumberger, 1981).

[6] For detailed discussion on other two methods refer to Leuven and Oosterbeek (2011).

This study uses data from the Periodic Labour Force Survey (PLFS) conducted by National Sample Survey Office (NSSO) for the period July 2018 to June 2019. The PLFS data provides the socio-economic and demographic information for a sample of 101,579 households (55,812 in rural areas and 45,767 in urban areas) comprising 420,757 individuals (rural – 239,817 and urban- 180,940) across all Indian states. The primary focus of the survey is to provide information on the economic activity status of all household members including their occupation and industry of work. This dataset also enables us to look at the status of workers from multiple perspectives such as enterprise type, location of work and social security benefits. The occupation codes are available at 3-digit level based on National Classification of Occupations (NCO) - 2004. Sample weights provided in the dataset are used to derive population estimates.

We restrict our analysis to the working age group, i.e., 15-59 years. This is to be consistent with the Government of India which considers age 15-59 years as economically active age group (NSSO, 2014). Besides age, PLFS also provides detailed data on the individual characteristics such as level of education, gender, place of residence, and so on along with the job-characteristics such as occupation, enterprise type, industry, and so on. Further, the survey provides information on the wages/salary (in cash and in-kind) earned by a worker. The main analysis in this article is conducted for wage/salary employed workers for whom information on both, education and occupation are available. Thus, the final sample comprises of 52,844 males and 15,060 females.

### 3.2. *Measurement: Informality and Education-Occupation Mismatch*

*Informality*

To measure the informality, we make use of three pieces of information provided in the PLFS: (i) *enterprise type* (proprietary, partnership, government/public sector, public/private limited company, and others); (ii) *number of workers* (less than ten, ten and above but less than 20, 20 and above, and not known); and (iii) *availability of social security benefits* (eligible and not eligible)[7]. Annexure 1 provides the summary statistics for these variables.

The first two that is, enterprise type and the number of workers ascertain whether a worker belongs to the formal or informal sector. Using NCEUS (2008) definition, the study identifies a worker to be from informal sector if she is working in either proprietary or a partnership firm that employs less than ten workers. Further, a worker is said to be informally employed if she is not eligible for any social security benefits. Besides, depending on both sector and employment, the study categorizes the workers into one of the four following categories: formally employed in the formal sector, formally employed in the informal sector, informally employed in the formal sector, and informally employed in the informal sector.

*Education-occupation mismatch*

Coming to the measurement of education-occupation mismatch (EOM), this article employs realized matches (RM) method, in particular, mean-based approach to

---

[7] There is one more option "not known" for the workers who do not know whether they receive any social security benefits. We exclude this category from our analysis. In this process, we lose 2,075 observations.

ascertain the match status of a worker.[8] To operationalise RM, we need two crucial information – years of education and occupation of a worker. The former is obtained by converting data on the level of general education into years of formal education following the conversion adapted from Duraisamy (2002).[9] To get the information on the latter, we use national classification of occupation (NCO) three-digit codes. There are 115 three-digit occupation titles[10] in our data. Using the information on education and occupation and sampling weights, we calculate the average years of formal education for every NCO 2004 three-digit code considering the wage/salary employed workers who are in the working-age group. Further, following Verdugo and Verdugo (1989), we establish a threshold of plus and minus one standard deviation from mean to ascertain the boundaries of required years of education for an occupation. In this scenario, a person would be overeducated if her education is more than mean plus one standard deviation, undereducated if her education is less than mean minus one standard deviation and adequately educated if her education falls in the bracket of the mean plus and minus one standard deviation. More precisely, suppose $e_i$ represents attained years of education of an individual '$i$', and $e_o$ and $s_o$ are the mean and standard deviation of years of education respectively for her occupation, $o$. As per RM, a worker will be considered:

Overeducated if: $e_i > e_o + s_o$;

---

[8] We do not use mode to avoid the problem of multi-mode in case of some occupations.

[9] No formal schooling corresponds to zero years of formal schooling, below primary corresponds to three years of formal education and completion of primary corresponds to five years of formal schooling. The corresponding years for middle, secondary, and higher secondary are eight, ten, and twelve respectively. Lastly, workers with graduate degree have been assigned 15 years of formal education and workers with post graduate have been assigned 17 years of formal education.

[10] NSSO (2014) embodies the full list of occupation list.

Undereducated if: $e_i < e_o - s_o$; and

Adequately educated if: $e_o - s_0 \leq e_i \leq e_o + s_o$

Using this definition, we find that the Indian labour market experiences a higher proportion of undereducation (16.07 percent) as compared to overeducation (14.63 percent).

Table 1 estimates the proportion of workers across the sector and employment. The percentage of workers in informal employment is higher than their formal counterparts while the opposite holds for informal sector workers.[11] The difference is more severe in the case of employment than sector. This indicates that segmentation of the labour market only by sector partially identifies the intensity of informality in India. Table 1 also provides information on the match status for the respective category of workers. Annexure 2 provides the daily average wages across the sector and employment.

<Table 1 here>

In our next section, we present an empirical model to estimate the differential impact of EOM on formal and informal workers after taking account of human capital variables, job characteristics, and other personal and regional characteristics.

## 4. Empirical Strategy

---

[11] The higher proportion of the workers in formal sector could be due the inconsideration of self-employed individuals in our analysis.

The seminal work of Becker (1964) and empirical modelling of Mincer (1974) is widely accepted as a standard model for estimating the returns to education and other human capital variables. The basic 'Mincerian' earnings function is:

$$log y_i = \alpha + \beta_1 S_i + \beta_2 Exp_i + \beta_3 Exp_i^2 + e_i \quad (1)$$

where the dependent variable is the logarithm of wages, $S$ represents years or level of education attained by an individual $i$, $Exp$ captures the experience, and $e$ is the disturbance term capturing the unobserved characteristics. The past studies have modified the above equation to control for job characteristics and other personal and household characteristics (Agrawal, 2012).

Therefore, equation (1) can be modified as follows:

$$log y_i = \alpha + \beta_1 S_i + \beta_2 Exp_i + \beta_3 Exp_i^2 + \beta_4 X_i + e_i \quad (1.1)$$

where $X$ represents the vector of other characteristics such as occupation, industry, and so on. Other variables are interpreted as before.

We can add a dummy for informality in equation (1.1) to take account of the sector of work and type of employment of a worker. Further, to capture the differential returns to EOM, Verdugo and Verdugo (1989) suggest adding the dummy for match status that is, overeducated, undereducated and adequately educated in equation (1.1). Hence, Verdugo and Verdugo's wage equation is as follows:

$$logy_i = \alpha + \delta_1 D_i^o + \delta_2 D_i^u + \delta_3 Inf_i + \beta_1 S_i + \beta_2 Exp_i + \beta_3 Exp_i^2 + \beta_4 X_i + e_i \quad (2)$$

where $D_i^o$ and $D_i^u$ are the dummy variables that take the value one if a person is over and under educated respectively and zero otherwise. $\delta_1$ and $\delta_2$ depict the returns to over and under educated respectively as compared to adequately educated. A positive (negative) $\delta_1$ will indicate that ceteris paribus, overeducated workers earn higher (lower) wages as compared to the adequately educated workers. Analogously, a positive (negative) $\delta_2$ will indicate that ceteris paribus, undereducated workers earn higher (lower) wages as compared to the adequately educated workers. $Inf$ represents the dummy for informality. Other variables are interpreted as before.

If the sample is randomly drawn from the population, we can estimate equation (2) to analyse the returns to wage-determining characteristics. However, only a subset of the population is employed and thus report wages. Another problem is that, wage information is provided solely for the wage/salaried workers making the data for self-employed unusable for the analysis. However, the choice of both, employment and wage-employment, is not random and thus, ignoring this dimension could lead to a sample selection bias. We use the Heckman selection model to account for both the decisions. This is done using a two-stage estimation procedure. The first step involves estimating the following participation equations:

$$Y_{1i}^* = z_{1i}\xi_1 + u_{1i} \quad (3.1)$$

$$Y_{2i}^* = z_{2i}\xi_2 + u_{2i} \quad (3.2)$$

where * indicates the unobserved variable. The dependent variable in equations (3.1) and (3.2) are binary variables where $Y_1$ takes the value one when a person is employed and zero otherwise, and $Y_2$ takes the value one when an individual is wage/salary employed and zero when she is self-employed. Further, $z_j, j = 1,2$ captures observed variables and $u_j, j = 1,2$ corresponds to the error term that captures the impact of unobserved variables. With the help of participation equations, we calculate inverse Mills ratios ($\lambda$), which is defined as the ratio of the probability density function to the cumulative distribution function. In the second step, the inverse Mills ratio is added as one of the independent variables in the wage equation to estimate unbiased coefficients.

Therefore, equation (2) is modified as follows to mitigate the sample selection bias.

$$logy_i = \alpha + \delta_1 D_i^o + \delta_2 D_i^u + \delta_3 Inf_i + \beta_1 S_i + \beta_2 Exp_i + \beta_3 Exp_i^2 + \beta_4 X_i + \lambda_{emp} + \lambda_{act} + e_i \quad (4)$$

where $\lambda_{emp}$ and $\lambda_{act}$ are inverse mills ratio obtained using participation equations (3.1) and (3.2) respectively. Other variables are interpreted as before. It is a prerequisite to identify at least one variable that does not affect the wages but influences the probability of participation (called as exclusion variables) to use this model.

In our study, we control for three set of covariates. The first group includes education level (no formal schooling, primary, middle, secondary, higher secondary, and graduates or above), age and its square. These variables are collectively named as human capital variables.[12] The second group comprises of job characteristics that includes occupation category (legislators, senior officials and managers, professionals, associate professionals, clerks, service workers and market sales workers, skilled agricultural and fishery workers, craft and related trades workers, plant and machine operators and assemblers, and elementary occupations), industry type (agriculture, manufacturing, construction, and services), the location of workplace (rural, urban, and no fixed location), and type of work contract (unwritten and written). The third group controls for a set of personal and household characteristics, namely, gender (male and female), marital status (unmarried, married, and others), the interaction of gender and marital status, social group (scheduled tribe, scheduled caste, other backward class and others), religion (Hindu, Muslim, Christian, and others), sector (rural and urban), and state of residence. These are the standard variables that are used in previous studies as well (Agrawal, 2012; Duraisamy, 2002). Therefore, the wage estimation equation is as follows:

$$log y_i = \alpha + \delta_1 D_i^o + \delta_2 D_i^u + \delta_3 Inf_i + \delta_4 HCV_i + \delta_5 JC_i + \delta_6 OC_i + \lambda_{emp} + \lambda_{act} + e_i \quad (5)$$

where outcome variable is the logarithm of daily wages and $HCV, JC, and\ OC$ are the vector of human capital variables, job characteristics, and other personal and household characteristics respectively. Other variables are interpreted as before.

---

[12] Generally, studies use experience or potential experience (age minus years of schooling minus five) instead of age. However, due to missing information on labour market experience in NSSO data and the unsuitably high proportion of workers with no formal schooling, both experience and potential experience are considered to be inadequate measures for the Indian data (Goel, 2017).

Further, in contrast to previous studies, we also control for the sector of work (formal and informal) while estimating equation (5) separately for formally and informally employed workers and vice-versa. Table 2 provides the summary statistics for these variables for the wage/salary employed working-age population (15-59 years).

<Table 2 HERE>

Coming to exclusion variables, we use number of dependent members (aged below six years or above 60 years) in a household, and household type and size as exclusion variables for equation (3.1). The choice of these variables is consistent with the literature that suggests that family characteristics variables are appropriate exclusion variables for the choice of work (Agrawal, 2012; Buchinsky, 2002). Further, we use dummy for having a hereditary vocational training as exclusion variables for equation (3.2). Therefore, we estimate participate equations (3.1) and (3.2) using respective exclusion variables, human capital variables, and personal and household characteristics. Note that we exclude job characteristics while estimating participation equations since these variables are only observed after an individual gets employed and hence cannot influence the probability of employment or wage/salary employment.

Based on the above empirical methodology, the next section discusses the results.

## 5. Results and Discussion

**Relative importance of informality and EOM**

As discussed earlier, from a worker's perspective, informality can affect her wages in two ways. First, informality can come from the firm's side. It means a firm is not registered and does not meet the threshold criterion to be considered eligible for formal registration. These firms due to small size face lot of difficulties such as credit constraints, labour attrition, lower productivity, and so on and therefore, are likely to pay less to their employees (Oi and Idson, 1999). The second effect of informality comes from nature of jobs offered to the workers. It is a well-known fact in developing countries that irrespective of nature of the firm, workers are provided employment without any social security, written job contracts, and other benefits. This leads to lower wages being offered due to the following reasons. Since the contracts are not written, workers do not have bargaining power in demanding higher wages. Also, there is an excess supply of workers which hinders the wage growth. Lastly, due to unwritten informal contracts, most of the labour legislation meant to protect workers remain out of the question. It is evident that these two dimensions of informality are structurally different and consequently their impact on wages of workers are likely to be different.

Table 3 provides the results for the informality while not taking account of EOM in Column 1-3.[13] Table 3 reveals that irrespective of the dimension of the informality discussed above, there is a penalty for workers in terms of lower wages (Column 1 and 2). Further, the penalty for being in informal employment (0.580) is more severe than that of being in the informal sector (0.187). This implies the cruciality of the type of employment over the sector of work. Additionally, the study explores the interplay of informality in both dimensions (Column 3).

---

[13] For the sake of brevity, the estimates of only concerned variables are given. The estimates for the full set of variables are available in the annexure.

<TABLE 3 HERE>

Based on that, there can be four categories of workers (sector – employment), formal – formal, formal – informal, informal -formal, and informal – informal. The corresponding proportion of workers in these categories is 39, 1, 32, and 28 respectively. As is clear from these numbers, the proportion of workers formally employed in the informal sector is minuscule, and hence point estimates should be interpreted with caution. It is observed that irrespective of the sector, informally employed workers receive lower wages. This again highlights the dominance of type of employment in determining the wages. One plausible reason for finding a higher formal-informal employment wage gap is as follows. In today's world, due to severe competition and to avoid strict labour regulations that usually results in higher costs, formal sector resorts to informalisation of jobs and thus create mixed-mode labour as a way out (Hussmanns, 2001). The author also argues that informalisation of jobs in the formal sector indicates that the informal sector as a concept has started to lose the relevance and need to be replaced by informal employment.

Another imperfection in the labour market can arise from a mismatch of education and occupation of workers which, in turn, affects the pecuniary benefits to the workers. Table 3 also analyses the impact of EOM without considering the informality in Column 4. It is found that undereducated workers do not receive a wage reward while overeducated workers earn a wage penalty of eight percent. The differential wages to match status (undereducated, overeducated, and adequately educated) arise from the fact that workers with the same level of education get varied wage premium due to their employment in different occupations. The overeducated worker, by definition, is working in an occupation that demands less education than her attained education and hence, represents the jobs that are less challenging and consequently lower paying concerning her education.

The results highlight the importance of both informality and EOM in determining the wages of workers. However, looking at the isolated literature on informality and EOM, one crucial observation that comes out is that these two kinds of literature do not account for each other (Herrera-Idárraga et al. (2015) being an exception). To understand this relationship, Table 4 provides the joint estimates of informality and EOM in Column 5 and 6.

The analysis reveals that including EOM does not exert any influence on the wage loss for informal workers. Consistent with the earlier results, they still suffer the penalty of the same magnitude. Further, the returns to EOM also remain consistent with the earlier findings after taking account of informality in the sector (Column 5). This represents that even though the sector of work matter for workers' wages, it is unrelated to their match status. However, the inclusion of informality in employment alters the magnitude of the penalty for being overeducated (Column 6). It resonates with the earlier results suggesting that the type of employment plays a more crucial role in determining the workers' wages. Next, Table 3 analyses the wage equation with EOM and four categories of workers based on the sector of work and type of employment (Column 7). It is found that the qualitative results are unchanged. Further, informally employed workers irrespective of the sector receive the highest penalty.

Thus, to summarize, both EOM and informality in employment are crucial in the labour market. However, informality in the sector is relatively less imperative for the wages when informalisation of jobs is taken into account. In a nutshell, it could be recommended to the policymakers that in the process of improving outcomes (mainly wages) for workers, the formalisation of jobs remains the primary factor. Some of the recent policies by the government of India are also directed in the same direction. One example being Pradhan Mantri Rojgar Protsahan Yojana (PMRPY) which is meant to facilitate social security benefits and consequently formalization of jobs by providing a partial subsidy on employee provident fund for social security provision (Naraparaju and Sharma, 2017).

The next part analyses the relevance of EOM for formal and informal workers.

**Relevance of EOM for formal and informal workers**

This sub-section deals with the question of whether EOM, a concept that originated in the context of developed countries, holds any significance in the labour market of developing countries where informality is predominant. For this, Table 4 estimates the rewards and penalty to mismatch across the dichotomy of formal and informal (sector of work and type of employment). We find that for informal sector workers, being mismatched does not seem to be relevant whereas overeducation is essential and crucial for workers in the formal sector (Column 1 and 2).[14] This finding indicates that matching of education with occupation plays a relatively less substantial role in the informal sector than the formal sector. The result could be owed to the characteristics of informal sector that is, jobs are unstructured, unorganized, and not well-defined.

---

[14] Note that these results are not comparable with the column 5 of the Table 3 that is a specification where dummies for both informality in sector and EOM are included. This is because in Table 4, while calculating returns for sector, we include a dummy for type of employment.

<TABLE 4 HERE>

Further, the analysis indicates that on segregating the data based on the type of employment, the central results related to EOM (undereducated workers are rewarded and overeducated workers are penalized in comparison to workers who match well with their respective occupations) remains inapplicable (Column 3 and 4 of Table 4). Further, the results highlight that while informally employed workers are not rewarded when they are undereducated, they are penalized in case of being overeducated (Column 3 of Table 4). This is worrisome since this makes informally employed workers the vulnerable group – no reward but a significant penalty.

## 6. Conclusion

This article investigates the intertwining relationship between informality (in terms of sector of work and type of employment) and education-occupation mismatch (EOM) in case of a developing economy. In particular, the article examines three aspects. First, the relative importance of informality and EOM in determining the workers' wages. Second, whether (and how much) the penalty or reward associated with EOM differ in the formal and informal economy. Third, whether (and how much) the formal-informal wage gap differ across workers with different match status.

The main results are as follows. First, the analysis finds that although both, informally employed workers and informal sector workers receive lower wages as compared to their formal counterparts, the wage difference is more severe in case of informal employment as compared to the informal sector. Second, the analysis finds that overeducated workers receive a wage penalty of eight percent as compared to adequately educated workers. Third, taking simultaneous account of both informality and EOM, the study reveals that inclusion of EOM does not exert any influence on the formal-informal wage gap. However, the inclusion of informality in employment alters the penalty for being mismatched. Fourth, considering the dichotomy of formal and informal reveals that EOM plays a significant role in determining wages for formally employed workers and formal sector workers. Further, while for informal sector workers, EOM is irrelevant, for informally employed workers, it is a double-edged sword. They do not get rewarded when they are undereducated but get a penalty on being overeducated as compared to their formal counterparts.

The key implications of this study are as follows. First, from the workers' perspective, it is not the sector of work but the type of employment that is crucial for higher returns to observable traits. This means that worker should be more concerned about the characteristics of the job (that is, whether it provides social security benefits, job security, fundamentally decent work) as compared to the nature of the job. From this study, the significant implication comes out is that EOM and informality are major concerns but informality is a more severe issue for developing economies. Another takeaway from this study is that the framework such as EOM which are largely applicable for developed countries should not be used off the shelf in the context of developing countries. As is clear from our study that in the formal part of the labour market, the EOM framework works similarly as in the developed countries, but in the informal part, it remains inapplicable.

To extend the understanding of this study one of the natural extensions is as follows. At a macro level, as labour market transitions from informal to formal, it would be interesting to observe what happens to the extent of EOM and its implications on workers' labour market outcomes. Further, how over time the relevance of informality and EOM would change in the labour market of developing economies can be another stimulating extension. This will facilitate and enhance the understanding of co-movement between EOM and informality.

# References

Agrawal, T. (2012). Returns to education in India: Some recent evidence. *Journal of Quantitative Economics*, *10*(2).

Becker, G. S. (1964). Human capital theory. *Columbia, New York*, *1964*.

Boeri, T., & Van Ours, J. (2013). *The economics of imperfect labor markets*. Princeton University Press.

Buchinsky, M. (2002). Quantile regression with sample selection: Estimating women's return to education in the US. *In Economic Applications of Quantile Regression* (pp. 87-113). Physica, Heidelberg.

Chevalier, A. (2003). Measuring over-education. *Economica*, 70(279), 509-531.

Duncan, G. J., & Hoffman, S. D. (1981). The incidence and wage effects of overeducation. *Economics of Education Review*, *1*(1), 75-86.

Duraisamy, P. (2002). Changes in returns to education in India, 1983–94: by gender, age-cohort and location. *Economics of Education Review*, *21*(6), 609-622.

Food and Agriculture Organization (FAO) (1998). "The state of food and agriculture 1998". Food and agriculture organization of the United Nations, Rome, 1998.

Fields, G. S. (1990). Labour market modelling and the urban informal sector: theory and evidence.

Freeman, R. (1976). The overeducated american.

Gindling, T. H. (1991). Labor market segmentation and the determination of wages in the public, private-formal, and informal sectors in San Jose, Costa Rica. *Economic Development and Cultural Change*, *39*(3), 585-605.

Goel, M. (2017). Inequality between and within skill groups: The curious case of India. *World Development*, 93, 153-176.

Harris, J. R., & Todaro, M. P. (1970). Migration, unemployment and development: a two-sector analysis. *The American Economic Review*, *60*(1), 126-142.

Hartog, J. (2000). Over-education and earnings: where are we, where should we go?. *Economics of Education Review*, 19(2), 131-147.

Herrera-Idárraga, P., López-Bazo, E., & Motellón, E. (2015). Double penalty in returns to education: informality and educational mismatch in the Colombian labour market. *The Journal of Development Studies*, *51*(12), 1683-1701.

Hien, P. V. (2018). Public investment in education and training in Vietnam. *International Education Studies*, *11*(7), 106-115.

Hussmanns, R. (2001, September). Informal sector and informal employment: elements of a conceptual framework. In *Fifth Meeting of the Expert Group on Informal Sector Statistics* (Delhi Group), New Delhi (pp. 19-21).

International Labour Office (ILO) (1998), "Report of the XVIth International Conference of Labour Statisticians", International Labour Organisation, Geneva.

International Labour Office (ILO) (2002): Decent Work and the Informal Economy; Report of the Director General; International Labour Conference, 90th Session; Report VI; International Labour Office, Geneva, 2002.

International Labour Office (ILO) (2003), "Guidelines Concerning a Statistical Definition of Informal Employment, Endorsed by the Seventeenth International Conference of Labour Statisticians" (24 November-3 December), Seventeenth International Conference of Labour Statisticians, International Labour Organisation, Geneva.

Kiker, B. F., Santos, M. C., & De Oliveira, M. M. (1997). Overeducation and undereducation: evidence for Portugal. *Economics of Education Review*, *16*(2), 111-125.

Leuven, E., & Oosterbeek, H. (2011). Overeducation and mismatch in the labor market. In *Handbook of the Economics of Education* (Vol. 4, pp. 283-326). Elsevier.

Lewis, W. A. (1954). Economic development with unlimited supplies of labour. The Manchester School, 22(2), 139-191.

McGuinness, S., Pouliakas, K., & Redmond, P. (2018). Skills mismatch: concepts, measurement and policy approaches. *Journal of Economic Surveys*.

Mincer, J. A. (1974). Schooling and earnings. In *Schooling, Experience, and Earnings* (pp. 41-63). NBER.

Naraparaju, K., & Sharma, A (2017). Labour regulations and worker welfare: The case of provident fund in India. *India Development Report 2017*.

National Commission for Enterprises in the Unorganised Sector (NCEUS) (2008), Report on Definitional and Statistical Issues Relating to Informal Economy, New Delhi, http://dcmsme.gov.in/Report_Statistical_Issues_Informal_Economy.pdf

National Sample Survey Office (NSSO) (2012). "Informal Sector and Conditions of Employment in India". Report No. 539 (66/10/2). New Delhi: Government of India.

National Sample Survey Office (NSSO) (2014). "Employment and Unemployment Situation in India". Report No. 554 (68/10/1). New Delhi: Government of India.

Nguyen, H. C., Nordman, C. J., & Roubaud, F. (2013). Who suffers the penalty?: A panel data analysis of earnings gaps in Vietnam. *Journal of Development Studies*, *49*(12), 1694-1710.

O'Higgins, N. (2003). Trends in the youth labour market in developing and transition countries.

Oi, W. Y., & Idson, T. L. (1999). Firm size and wages. *Handbook of Labor Economics*, *3*, 2165-2214.

Pradhan, M., & Van Soest, A. (1995). Formal and informal sector employment in urban areas of Bolivia. *Labour economics*, *2*(3), 275-297.

Quinn, M. A., & Rubb, S. (2006). Mexico's labor market: The importance of education-occupation matching on wages and productivity in developing countries. *Economics of Education Review*, *25*(2), 147-156.

Ramaswamy, K. V. (2013). Wages of unorganized sector workers in India: What NSS survey data can tell us?. *Indian Journal of Industrial Relations*, 438-449.

Rumberger, R. W. (1981). The rising incidence of overeducation in the US labor market. *Economics of Education Review*, *1*(3), 293-314.

Ruppert Bulmer, E. (2018). Defining informality vs mitigating its negative effects. *IZA World of Labor*.

Sattinger, M. (1993). Assignment models of the distribution of earnings. *Journal of Economic Literature*, *31*(2), 831-880.

Thurow, L. C. (1975). *Generating inequality*. Basic books.

Unni, J., & Sarkar, S. (2013). Wages and informal labour markets In India: Whither premiums on human capital investments?. *Indian Journal of Labour Economics*, *56*(4).

Verdugo, R. R., & Verdugo, N. T. (1989). The impact of surplus schooling on earnings: Some additional findings. *Journal of Human Resources*, 629-643.

Verhaest, D., & Omey, E. (2006). The impact of overeducation and its measurement. *Social Indicators Research*, *77*(3), 419-448.

# Tables

Table 1: Proportion of worker across sector and employment: by match status

| Job Type | | Overall | Match Status | | |
|---|---|---|---|---|---|
| | | | Under | Adequately | Over |
| Sector | Formal | 62 | 18 | 68 | 14 |
| | Informal | 38 | 23 | 64 | 13 |
| Employment | Formal | 29 | 10 | 73 | 17 |
| | Informal | 71 | 23 | 64 | 13 |
| Sector-Employment | Inf-Inf | 39 | 23 | 64 | 13 |
| | Inf-For | 1 | 16 | 63 | 21 |
| | For-Inf | 32 | 23 | 64 | 13 |
| | For-For | 28 | 10 | 74 | 16 |

*Source*: Authors' calculation based on Periodic Labour Force Survey, 2018-19.
*Note:* Sampling weights have been used.

Table 2: Summary statistics for wage/salary employed in working-age group: by sector and employment

| Variables | | Overall | Sector | | Employment | |
|---|---|---|---|---|---|---|
| | | | Formal | Informal | Formal | Informal |
| ***Education*** | No formal schooling | 0.14 | 0.12 | 0.21 | 0.03 | 0.21 |
| | Primary | 0.15 | 0.14 | 0.27 | 0.05 | 0.26 |
| | Middle | 0.21 | 0.14 | 0.23 | 0.09 | 0.21 |
| | Secondary | 0.15 | 0.14 | 0.14 | 0.14 | 0.14 |
| | Higher Secondary | 0.14 | 0.16 | 0.08 | 0.22 | 0.09 |
| | Graduate or above | 0.2 | 0.3 | 0.06 | 0.46 | 0.09 |
| ***Age*** | | 36.57 | 37.48 | 33.19 | 40.94 | 33.46 |
| **Gender** | Male | 0.78 | 0.79 | 0.89 | 0.82 | 0.83 |
| | Female | 0.22 | 0.21 | 0.11 | 0.18 | 0.17 |
| **Marital status** | Never married | 0.22 | 0.18 | 0.27 | 0.11 | 0.26 |
| | Currently married | 0.73 | 0.78 | 0.69 | 0.85 | 0.69 |
| | Others | 0.05 | 0.05 | 0.03 | 0.04 | 0.05 |
| **Social group** | Scheduled tribe | 0.13 | 0.15 | 0.09 | 0.19 | 0.11 |
| | Scheduled caste | 0.22 | 0.17 | 0.23 | 0.12 | 0.23 |
| | Other backward class | 0.38 | 0.33 | 0.43 | 0.3 | 0.4 |
| | Others | 0.28 | 0.34 | 0.25 | 0.39 | 0.27 |
| **Religion** | Hinduism | 0.76 | 0.76 | 0.75 | 0.74 | 0.77 |
| | Islam | 0.12 | 0.11 | 0.17 | 0.09 | 0.14 |
| | Christianity | 0.07 | 0.09 | 0.04 | 0.12 | 0.05 |
| | Others | 0.05 | 0.04 | 0.04 | 0.05 | 0.04 |
| **Sector** | Rural | 0.49 | 0.47 | 0.52 | 0.42 | 0.52 |
| | Urban | 0.51 | 0.53 | 0.48 | 0.58 | 0.48 |
| **Occupation** | Legislators, Senior Officials and Managers | 0.03 | 0.03 | 0.01 | 0.05 | 0.01 |

| Variables | | Overall | Sector | | Employment | |
|---|---|---|---|---|---|---|
| | | | Formal | Informal | Formal | Informal |
| | Professionals | 0.08 | 0.12 | 0.02 | 0.19 | 0.04 |
| | Associate Professionals | 0.09 | 0.18 | 0.04 | 0.27 | 0.06 |
| | Clerks | 0.06 | 0.1 | 0.02 | 0.14 | 0.03 |
| | Service Workers and Shop & Market Sales Workers | 0.13 | 0.11 | 0.14 | 0.13 | 0.12 |
| | Skilled Agricultural and Fishery Workers | 0.02 | 0.01 | 0.01 | 0.01 | 0.01 |
| | Craft and Related Trades Workers | 0.16 | 0.15 | 0.3 | 0.07 | 0.27 |
| | Plant and Machine Operators and Assemblers | 0.09 | 0.08 | 0.12 | 0.07 | 0.11 |
| | Elementary Occupations | 0.34 | 0.22 | 0.34 | 0.07 | 0.36 |
| **Location of Work** | Rural | 0.32 | 0.43 | 0.47 | 0.39 | 0.47 |
| | Urban | 0.62 | 0.55 | 0.48 | 0.61 | 0.49 |
| | No-fixed location | 0.06 | 0.02 | 0.05 | 0 | 0.04 |
| **Type of job contract** | Unwritten | 0.74 | 0.58 | 0.97 | 0.32 | 0.92 |
| | Written | 0.26 | 0.42 | 0.03 | 0.68 | 0.08 |

*Source*: Authors' calculation based on Periodic Labour Force Survey, 2018-19.

Table 3: Returns to informality and EOM

| Explanatory Variables | (1) | (2) | (3) | (4) | (5) | (6) | (7) |
|---|---|---|---|---|---|---|---|
| **Sector (Base Cat.: Formal)** | | | | | | | |
| Informal | -0.187*** | | | | -0.188*** | | |
|  | (0.00833) | | | | (0.00831) | | |
| **Employment (Base Cat.: Formal)** | | | | | | | |
| Informal | | -0.580*** | | | | -0.578*** | |
|  | | (0.00954) | | | | (0.00951) | |
| **Sector-Employment (Base Cat.: Formal-Formal)** | | | | | | | |
| Informal-Informal | | | -0.599*** | | | | -0.599*** |
|  | | | (0.0112) | | | | (0.0112) |
| Informal-Formal | | | -0.362*** | | | | -0.360*** |
|  | | | (0.0278) | | | | (0.0278) |
| Formal-Informal | | | -0.599*** | | | | -0.598*** |
|  | | | (0.0100) | | | | (0.00996) |
| **Match Status (Base Cat.: Adequately Educated)** | | | | | | | |
| Undereducated | | | | -0.00546 | -0.00235 | -0.0164 | -0.0149 |
|  | | | | (0.0127) | (0.0126) | (0.0122) | (0.0121) |
| Overeducated | | | | -0.0804*** | -0.0814*** | -0.0615*** | -0.0606*** |
|  | | | | (0.0108) | (0.0107) | (0.0102) | (0.0102) |
| Number of Observations | 40,974 | 38,459 | 38,455 | 40,961 | 40,957 | 38,442 | 38,438 |
| R-squared | 0.500 | 0.573 | 0.576 | 0.494 | 0.501 | 0.574 | 0.577 |

*Source*: Authors' calculation based on Periodic Labour Force Survey, 2018-19.
*Note*: (i) *** signals significant at 1% level, ** signals significant at 5% level, and * signals significant at 10% level.
(ii) Robust standard errors are in parenthesis.

Table 4: Returns to EOM: by sector of work and type of employment

| Explanatory Variables | (1) | (2) | (3) | (4) |
|---|---|---|---|---|
| | Sector | | Employment | |
| | Informal | Formal | Informal | Formal |
| **Match Status (Base Cat.: Adequately Educated)** | | | | |
| Undereducated | -0.0113 | 0.00320 | 0.00501 | 0.00665 |
| | (0.0189) | (0.0146) | (0.0149) | (0.0182) |
| Overeducated | -0.00350 | -0.0682*** | -0.0319** | -0.0525*** |
| | (0.0184) | (0.0117) | (0.0129) | (0.0139) |
| ***Observations*** | 9,295 | 29,143 | 19,702 | 18,736 |
| ***R-Squared*** | 0.350 | 0.569 | 0.365 | 0.419 |

*Source*: Authors' calculation based on Periodic Labour Force Survey, 2018-19.
*Note*: (i) *** signals significant at 1% level, ** signals significant at 5% level, and * signals significant at 10% level.
    (ii) Robust standard errors are in parenthesis.

# Annexure

Table A1: Summary statistics: Enterprise type, Number of workers, Social security benefits

|  | Proportion of Workers (in %) |
|---|---|
| ***Enterprise Type*** | |
| Proprietary | 51.72 |
| Partnership | 1.40 |
| Government/Public Sector | 19.74 |
| Public/Private Limited Company | 15.76 |
| Others | 11.38 |
| ***Number of Workers*** | |
| Less than 10 | 57.69 |
| 10-19 | 10.40 |
| Above 19 | 25.50 |
| Not Known | 6.41 |
| ***Social Security Benefits*** | |
| Eligible | 28.89 |
| Not Eligible | 71.11 |

*Source*: Authors' calculation based on Periodic Labour Force Survey, 2018-19.
*Note:* Sampling weights have been used.

Table A2: Daily average wages (in Indian rupees): by sector and employment

| Job Type | | Overall | Match Status | | |
|---|---|---|---|---|---|
| | | | Under | Adequately | Over |
| Sector | Formal | 631 | 435 | 668 | 646 |
| | Informal | 312 | 282 | 313 | 349 |
| Employment | Formal | 862 | 696 | 898 | 805 |
| | Informal | 314 | 260 | 322 | 351 |
| Sector-Employment | Inf-Inf | 296 | 268 | 299 | 322 |
| | Inf-For | 533 | 434 | 538 | 590 |
| | For-Inf | 327 | 255 | 340 | 373 |
| | For-For | 875 | 709 | 911 | 813 |

*Source*: Authors' calculation based on Periodic Labour Force Survey, 2018-19.
*Note:* Sampling weights have been used.

|  | Employment | Type of Employment |
|---|---|---|
| **Table A3: Sample Selection Probit Model – in working-age group** | | |
| ***Education (Base Cat.: No Formal Schooling)*** | | |
| Primary | 0.116*** | 0.116*** |
|  | (0.0110) | (0.0128) |
| Middle | 0.0547*** | 0.176*** |
|  | (0.0106) | (0.0125) |
| Secondary | -0.0652*** | 0.141*** |
|  | (0.0117) | (0.0136) |
| Higher Secondary | -0.179*** | 0.0266* |
|  | (0.0125) | (0.0145) |
| Graduate or Above | 0.150*** | -0.340*** |
|  | (0.0133) | (0.0145) |
| **Age** | 0.279*** | 0.0226*** |
|  | (0.00219) | (0.00276) |
| **Age squared** | -0.00333*** | -0.000111*** |
|  | (2.80e-05) | (3.44e-05) |
| ***Gender (Base Cat.: Male)*** | | |
| Female | -2.090*** | 0.0832*** |
|  | (0.00898) | (0.00923) |
| ***Marital Status (Base Cat.: Unmarried)*** | | |
| Married | 0.397*** | 0.115*** |
|  | (0.0127) | (0.0125) |
| Others | 0.829*** | -0.216*** |
|  | (0.0214) | (0.0221) |
| ***Social Group (Base Cat.: Scheduled Tribe)*** | | |
| Scheduled Caste | -0.0948*** | -0.364*** |
|  | (0.0138) | (0.0152) |
| OBC | -0.118*** | 0.0825*** |
|  | (0.0127) | (0.0138) |
| Others | -0.216*** | 0.163*** |
|  | (0.0131) | (0.0145) |
| ***Religion (Base Cat.: Hindu)*** | | |
| Muslim | -0.0424*** | 0.0449*** |
|  | (0.00984) | (0.0122) |
| Christian | 0.0143 | -0.0247 |
|  | (0.0184) | (0.0202) |
| Others | -0.0116 | 0.0241 |
|  | (0.0178) | (0.0201) |
| ***Sector (Base Cat.: Rural)*** | | |
| Urban | -0.155*** | -0.406*** |
|  | (0.00696) | (0.00795) |
| ***Exclusion Variables*** | | |
| Dependent | 0.0571*** | |

|  | Employment | Type of Employment |
|---|---|---|
|  | (0.00337) |  |
| Household Type (Base Cat.: Self-employed) |  |  |
| regular wage/salary earning | 0.0236*** |  |
|  | (0.00756) |  |
| casual labour | 0.0843*** |  |
|  | (0.00903) |  |
| others | -1.347*** |  |
|  | (0.0225) |  |
| Household Size | -0.0486*** |  |
|  | (0.00227) |  |
| Vocational Education (Base Cat.: Yes) |  |  |
| No Vocational Education |  | 0.883*** |
|  |  | (0.0235) |

*Source*: Authors' calculation based on Periodic Labour Force Survey, 2018-19.
*Note*: (i) *** signals significant at 1% level, ** signals significant at 5% level, and * signals significant at 10% level.
 (ii) Robust standard errors are in parenthesis.
 (iii) The analysis also controls for broad industry groups, interaction of gender and marital status, and 35 states and union territories.

| Explanatory Variables | (1) | (2) | (3) | (4) | (5) | (6) | (7) |
|---|---|---|---|---|---|---|---|
| **Table A4: Returns to informality and EOM** | | | | | | | |
| ***Sector (Base Cat.: Formal)*** | | | | | | | |
| Informal | -0.187*** | | | | -0.188*** | | |
|  | (0.00833) | | | | (0.00831) | | |
| ***Employment (Base Cat.: Formal)*** | | | | | | | |
| Informal | | -0.580*** | | | | -0.578*** | |
|  | | (0.00954) | | | | (0.00951) | |
| ***Sector-Employment (Base Cat.: Formal-Formal)*** | | | | | | | |
| Informal-Informal | | | -0.599*** | | | | -0.599*** |
|  | | | (0.0112) | | | | (0.0112) |
| Informal-Formal | | | -0.362*** | | | | -0.360*** |
|  | | | (0.0278) | | | | (0.0278) |
| Formal-Informal | | | -0.599*** | | | | -0.598*** |
|  | | | (0.0100) | | | | (0.00996) |
| ***Match Status (Base Cat.: Adequately Educated)*** | | | | | | | |
| Undereducated | | | | -0.00546 | -0.00235 | -0.0164 | -0.0149 |
|  | | | | (0.0127) | (0.0126) | (0.0122) | (0.0121) |
| Overeducated | | | | -0.0804*** | -0.0814*** | -0.0615*** | -0.0606*** |
|  | | | | (0.0108) | (0.0107) | (0.0102) | (0.0102) |
| ***Education (Base Cat.: No Formal Schooling)*** | | | | | | | |
| Primary | 0.139*** | 0.110*** | 0.110*** | 0.134*** | 0.141*** | 0.101*** | 0.102*** |
|  | (0.0174) | (0.0170) | (0.0170) | (0.0194) | (0.0195) | (0.0190) | (0.0189) |
| Middle | 0.238*** | 0.176*** | 0.177*** | 0.243*** | 0.242*** | 0.166*** | 0.168*** |
|  | (0.0176) | (0.0174) | (0.0173) | (0.0212) | (0.0212) | (0.0207) | (0.0207) |
| Secondary | 0.308*** | 0.228*** | 0.228*** | 0.326*** | 0.321*** | 0.225*** | 0.226*** |
|  | (0.0173) | (0.0171) | (0.0170) | (0.0209) | (0.0209) | (0.0206) | (0.0205) |

| Explanatory Variables | (1) | (2) | (3) | (4) | (5) | (6) | (7) |
|---|---|---|---|---|---|---|---|
| Higher Secondary | 0.448*** | 0.343*** | 0.340*** | 0.491*** | 0.480*** | 0.352*** | 0.351*** |
|  | (0.0163) | (0.0160) | (0.0160) | (0.0223) | (0.0222) | (0.0217) | (0.0216) |
| Graduate or Above | 0.715*** | 0.584*** | 0.579*** | 0.764*** | 0.762*** | 0.600*** | 0.596*** |
|  | (0.0252) | (0.0249) | (0.0249) | (0.0317) | (0.0314) | (0.0307) | (0.0307) |
| **Age** | 0.00604 | 0.00414 | 0.00476 | 0.00739* | 0.00600 | 0.00399 | 0.00459 |
|  | (0.00375) | (0.00371) | (0.00370) | (0.00377) | (0.00374) | (0.00370) | (0.00369) |
| **Age Squared** | 0.000108** | 8.88e-05** | 7.99e-05* | 0.000104** | 0.000109** | 9.19e-05** | 8.30e-05* |
|  | (4.43e-05) | (4.37e-05) | (4.35e-05) | (4.46e-05) | (4.42e-05) | (4.36e-05) | (4.34e-05) |
| **Occupation (Base Cat.: Elementary Occupation)** | | | | | | | |
| Legislators, Senior Officials and Managers | 0.485*** | 0.418*** | 0.419*** | 0.430*** | 0.437*** | 0.386*** | 0.388*** |
|  | (0.0204) | (0.0197) | (0.0196) | (0.0228) | (0.0227) | (0.0219) | (0.0218) |
| Professionals | 0.393*** | 0.349*** | 0.350*** | 0.341*** | 0.346*** | 0.319*** | 0.320*** |
|  | (0.0160) | (0.0153) | (0.0153) | (0.0187) | (0.0186) | (0.0178) | (0.0177) |
| Associate Professionals | 0.240*** | 0.195*** | 0.196*** | 0.196*** | 0.203*** | 0.172*** | 0.173*** |
|  | (0.0151) | (0.0147) | (0.0147) | (0.0174) | (0.0172) | (0.0166) | (0.0166) |
| Clerks | 0.210*** | 0.156*** | 0.155*** | 0.170*** | 0.178*** | 0.137*** | 0.136*** |
|  | (0.0146) | (0.0141) | (0.0142) | (0.0166) | (0.0165) | (0.0159) | (0.0159) |
| Service Workers and Shop & Market Sales Workers | 0.0911*** | 0.0748*** | 0.0778*** | 0.0460*** | 0.0804*** | 0.0675*** | 0.0707*** |
|  | (0.0121) | (0.0119) | (0.0121) | (0.0122) | (0.0123) | (0.0121) | (0.0122) |
| Skilled Agricultural and Fishery Workers | 0.0176 | 0.0220 | 0.0281 | -0.00585 | 0.0196 | 0.0222 | 0.0284 |
|  | (0.0371) | (0.0335) | (0.0334) | (0.0380) | (0.0371) | (0.0335) | (0.0334) |
| Craft and Related Trades Workers | 0.143*** | 0.142*** | 0.145*** | 0.0985*** | 0.136*** | 0.137*** | 0.141*** |
|  | (0.0125) | (0.0128) | (0.0128) | (0.0126) | (0.0126) | (0.0129) | (0.0129) |
| Plant and Machine Operators and Assemblers | 0.192*** | 0.180*** | 0.181*** | 0.159*** | 0.187*** | 0.178*** | 0.179*** |
|  | (0.0123) | (0.0127) | (0.0128) | (0.0122) | (0.0122) | (0.0126) | (0.0126) |

| Explanatory Variables | (1) | (2) | (3) | (4) | (5) | (6) | (7) |
|---|---|---|---|---|---|---|---|
| **Location of Workplace (Base Cat.: Rural)** | | | | | | | |
| Urban | 0.0842*** | 0.0686*** | 0.0683*** | 0.0782*** | 0.0851*** | 0.0695*** | 0.0692*** |
|  | (0.0113) | (0.0111) | (0.0111) | (0.0111) | (0.0110) | (0.0109) | (0.0109) |
| No Fixed Location | 0.140*** | 0.161*** | 0.164*** | 0.0785*** | 0.142*** | 0.161*** | 0.164*** |
|  | (0.0208) | (0.0213) | (0.0213) | (0.0202) | (0.0208) | (0.0212) | (0.0213) |
| **Contract (Base Cat.: Unwritten)** | | | | | | | |
| Written | 0.387*** | 0.177*** | 0.167*** | 0.434*** | 0.387*** | 0.178*** | 0.168*** |
|  | (0.0103) | (0.0104) | (0.0104) | (0.00988) | (0.0102) | (0.0102) | (0.0103) |
| **Gender (Base Cat.: Male)** | | | | | | | |
| Female | -0.463*** | -0.400*** | -0.402*** | -0.433*** | -0.460*** | -0.397*** | -0.399*** |
|  | (0.0243) | (0.0241) | (0.0240) | (0.0244) | (0.0243) | (0.0240) | (0.0240) |
| **Marital Status (Base Cat.: Unmarried)** | | | | | | | |
| Married | 0.108*** | 0.0834*** | 0.0837*** | 0.116*** | 0.106*** | 0.0822*** | 0.0825*** |
|  | (0.0117) | (0.0114) | (0.0114) | (0.0118) | (0.0117) | (0.0115) | (0.0114) |
| Others | 0.0829*** | 0.0828*** | 0.0812*** | 0.0779*** | 0.0799*** | 0.0801*** | 0.0786*** |
|  | (0.0242) | (0.0231) | (0.0231) | (0.0242) | (0.0242) | (0.0231) | (0.0231) |
| **Social Group (Base Cat.: Scheduled Tribe)** | | | | | | | |
| Scheduled Caste | -0.0851*** | -0.0497* | -0.0541** | -0.0990*** | -0.0871*** | -0.0518** | -0.0561** |
|  | (0.0259) | (0.0254) | (0.0254) | (0.0259) | (0.0259) | (0.0253) | (0.0254) |
| OBC | -0.0960*** | -0.0744*** | -0.0771*** | -0.0979*** | -0.0952*** | -0.0735*** | -0.0761*** |
|  | (0.0163) | (0.0156) | (0.0156) | (0.0164) | (0.0163) | (0.0156) | (0.0155) |
| Others | -0.000403 | 0.00759 | 0.00643 | 0.00530 | 0.000999 | 0.00879 | 0.00758 |
|  | (0.0177) | (0.0170) | (0.0170) | (0.0178) | (0.0177) | (0.0170) | (0.0170) |
| **Religion (Base Cat.: Hindu)** | | | | | | | |
| Muslim | -0.0208* | -0.0118 | -0.0109 | -0.0310*** | -0.0213* | -0.0123 | -0.0114 |
|  | (0.0113) | (0.0110) | (0.0110) | (0.0114) | (0.0112) | (0.0110) | (0.0109) |
| Christian | 0.0206 | 0.0160 | 0.0125 | 0.0225 | 0.0209 | 0.0163 | 0.0128 |

| Explanatory Variables | (1) | (2) | (3) | (4) | (5) | (6) | (7) |
|---|---|---|---|---|---|---|---|
| | (0.0188) | (0.0171) | (0.0171) | (0.0190) | (0.0188) | (0.0171) | (0.0171) |
| Others | 0.0173 | 0.0188 | 0.0207 | 0.0101 | 0.0180 | 0.0192 | 0.0212 |
| | (0.0173) | (0.0167) | (0.0165) | (0.0175) | (0.0174) | (0.0167) | (0.0165) |
| **Sector (Base Cat.: Rural)** | | | | | | | |
| Urban | 0.148*** | 0.155*** | 0.152*** | 0.139*** | 0.144*** | 0.151*** | 0.148*** |
| | (0.0231) | (0.0231) | (0.0231) | (0.0231) | (0.0230) | (0.0230) | (0.0231) |
| **Selection term** | | | | | | | |
| Work | -0.0147 | -0.0287 | -0.0263 | -0.0255 | -0.0174 | -0.0314 | -0.0290 |
| | (0.0201) | (0.0201) | (0.0200) | (0.0202) | (0.0201) | (0.0200) | (0.0200) |
| Wage/salaried Employment | -0.0512 | -0.127* | -0.121 | 0.0101 | -0.0418 | -0.117 | -0.112 |
| | (0.0746) | (0.0747) | (0.0749) | (0.0750) | (0.0745) | (0.0746) | (0.0748) |
| Number of Observations | 40,974 | 38,459 | 38,455 | 40,961 | 40,957 | 38,442 | 38,438 |
| R-squared | 0.500 | 0.573 | 0.576 | 0.494 | 0.501 | 0.574 | 0.577 |

*Source*: Authors' calculation based on Periodic Labour Force Survey, 2018-19.
*Note*: (i) *** signals significant at 1% level, ** signals significant at 5% level, and * signals significant at 10% level.
    (ii) Robust standard errors are in parenthesis.
    (iii) The analysis also controls for broad industry groups, interaction of gender and marital status, and 35 states and union territories.

| Explanatory Variables | 1 | 2 | 3 | 4 |
|---|---|---|---|---|
| | Sector | | Employment | |
| | Informal | Formal | Informal | Formal |
| **Match Status (Base Cat.: Adequately Educated)** | | | | |
| Undereducated | -0.0113 | 0.00320 | 0.00501 | 0.00665 |
| | (0.0189) | (0.0146) | (0.0149) | (0.0182) |
| Overeducated | -0.00350 | -0.0682*** | -0.0319** | -0.0525*** |
| | (0.0184) | (0.0117) | (0.0129) | (0.0139) |
| **Education (Base Cat.: No Formal Schooling)** | | | | |
| Primary | 0.0858*** | 0.103*** | 0.0993*** | 0.0482 |
| | (0.0263) | (0.0249) | (0.0199) | (0.0427) |
| Middle | 0.124*** | 0.193*** | 0.156*** | 0.147*** |
| | (0.0307) | (0.0265) | (0.0227) | (0.0433) |
| Secondary | 0.137*** | 0.276*** | 0.201*** | 0.226*** |
| | (0.0310) | (0.0262) | (0.0230) | (0.0426) |
| Higher Secondary | 0.207*** | 0.427*** | 0.268*** | 0.407*** |
| | (0.0326) | (0.0273) | (0.0246) | (0.0428) |
| Graduate or Above | 0.331*** | 0.698*** | 0.471*** | 0.580*** |
| | (0.0461) | (0.0393) | (0.0355) | (0.0580) |
| Age | 0.0434*** | -0.000677 | 0.0417*** | 0.0171*** |
| | (0.00580) | (0.00455) | (0.00446) | (0.00592) |
| Age Squared | -0.000520*** | 0.000168*** | -0.000492*** | 4.26e-05 |
| | (7.19e-05) | (5.22e-05) | (5.41e-05) | (6.55e-05) |
| **Occupation (Base Cat.: Elementary Occupation)** | | | | |
| Legislators, Senior Officials and Managers | 0.349*** | 0.362*** | 0.364*** | 0.437*** |
| | (0.0450) | (0.0240) | (0.0329) | (0.0297) |
| Professionals | 0.195*** | 0.304*** | 0.181*** | 0.404*** |

| Explanatory Variables | 1 | 2 | 3 | 4 |
|---|---|---|---|---|
| | Sector | | Employment | |
| | Informal | Formal | Informal | Formal |
| | (0.0378) | (0.0200) | (0.0237) | (0.0263) |
| Associate Professionals | 0.0802** | 0.163*** | 0.0669*** | 0.255*** |
| | (0.0336) | (0.0188) | (0.0206) | (0.0252) |
| Clerks | 0.167*** | 0.112*** | 0.122*** | 0.168*** |
| | (0.0318) | (0.0181) | (0.0211) | (0.0237) |
| Service Workers and Shop & Market Sales Workers | 0.0687*** | 0.0716*** | 0.0412*** | 0.169*** |
| | (0.0197) | (0.0150) | (0.0134) | (0.0221) |
| Skilled Agricultural and Fishery Workers | 0.0498 | 0.0237 | 0.00776 | 0.0316 |
| | (0.0691) | (0.0363) | (0.0419) | (0.0455) |
| Craft and Related Trades Workers | 0.105*** | 0.120*** | 0.0837*** | 0.147*** |
| | (0.0213) | (0.0155) | (0.0141) | (0.0217) |
| Plant and Machine Operators and Assemblers | 0.209*** | 0.139*** | 0.189*** | 0.126*** |
| | (0.0208) | (0.0153) | (0.0137) | (0.0215) |
| **Location of Workplace (Base Cat.: Rural)** | | | | |
| Urban | 0.0347** | 0.0718*** | 0.0917*** | 0.0407*** |
| | (0.0162) | (0.0134) | (0.0130) | (0.0150) |
| No Fixed Location | 0.107*** | 0.174*** | 0.163*** | -0.117 |
| | (0.0232) | (0.0495) | (0.0218) | (0.0801) |
| **Contract (Base Cat.: Unwritten)** | | | | |
| Written | 0.151*** | 0.159*** | 0.102*** | 0.169*** |
| | (0.0290) | (0.0109) | (0.0148) | (0.0135) |
| **Gender (Base Cat.: Male)** | | | | |
| Female | -0.503*** | -0.392*** | -0.618*** | -0.244*** |
| | (0.0444) | (0.0279) | (0.0302) | (0.0370) |
| **Marital Status (Base Cat.: Unmarried)** | | | | |
| Married | 0.120*** | 0.0639*** | 0.0972*** | 0.0786*** |

| Explanatory Variables | 1 | 2 | 3 | 4 |
|---|---|---|---|---|
| | Sector | | Employment | |
| | Informal | Formal | Informal | Formal |
| | (0.0164) | (0.0146) | (0.0130) | (0.0198) |
| Others | 0.0978** | 0.0784*** | 0.127*** | 0.0415 |
| | (0.0405) | (0.0277) | (0.0277) | (0.0374) |
| **Social Group (Base Cat.: Scheduled Tribe)** | | | | |
| Scheduled Caste | -0.00289 | -0.0591* | -0.0172 | -0.157*** |
| | (0.0349) | (0.0328) | (0.0288) | (0.0443) |
| OBC | -0.0137 | -0.0924*** | -0.0157 | -0.0936*** |
| | (0.0258) | (0.0182) | (0.0200) | (0.0226) |
| Others | 0.0744*** | -0.0147 | 0.0762*** | -0.00555 |
| | (0.0283) | (0.0204) | (0.0220) | (0.0258) |
| **Religion (Base Cat.: Hindu)** | | | | |
| Muslim | -0.0151 | -0.0216 | -0.0262** | -0.0295* |
| | (0.0153) | (0.0138) | (0.0125) | (0.0178) |
| Christian | -0.0378 | 0.0196 | 0.0313 | -0.00757 |
| | (0.0340) | (0.0188) | (0.0252) | (0.0216) |
| Others | 0.0253 | 0.0107 | 0.0300 | 0.00872 |
| | (0.0253) | (0.0202) | (0.0219) | (0.0254) |
| **Sector (Base Cat.: Rural)** | | | | |
| Urban | 0.0354 | 0.203*** | 0.0667*** | 0.136*** |
| | (0.0281) | (0.0312) | (0.0249) | (0.0425) |
| **Selection term** | | | | |
| Work | 0.0256 | -0.0257 | 0.0625** | -0.0264 |
| | (0.0341) | (0.0237) | (0.0243) | (0.0311) |
| Wage/salaried Employment | -0.0742 | -0.155 | -0.0586 | 0.151 |
| | (0.0896) | (0.102) | (0.0796) | (0.143) |
| Number of Observations | 9,295 | 29,143 | 19,702 | 18,736 |

| Explanatory Variables | 1 | 2 | 3 | 4 |
|---|---|---|---|---|
| | Sector | | Employment | |
| | Informal | Formal | Informal | Formal |
| R-squared | 0.350 | 0.569 | 0.365 | 0.419 |

*Source*: Authors' calculation based on Periodic Labour Force Survey, 2018-19.
*Note*: (i) *** signals significant at 1% level, ** signals significant at 5% level, and * signals significant at 10% level.
  (ii) Robust standard errors are in parenthesis.
    (iii) The analysis also controls for broad industry groups, interaction of gender and marital status, and 35 states and union territories.

Table A6: Formal-informal employment wage gap: by match status

| Explanatory Variables | (1) Under | (2) Adequate | (3) Over |
|---|---|---|---|
| **Employment (Base Cat.: Formal)** | | | |
| Informal | -0.593*** | -0.578*** | -0.532*** |
|  | (0.0211) | (0.00852) | (0.0172) |
| **Education (Base Cat.: No Formal Schooling)** | | | |
| Primary | 0.0760*** | 0.0951 | - |
|  | (0.0273) | (0.0977) |  |
| Middle | 0.124*** | 0.177* | -0.240*** |
|  | (0.0445) | (0.0977) | (0.0868) |
| Secondary | 0.307*** | 0.217** | -0.227*** |
|  | (0.0440) | (0.0977) | (0.0622) |
| Higher Secondary | 0.484*** | 0.366*** | -0.135*** |
|  | (0.0444) | (0.0978) | (0.0459) |
| Graduate or Above | 0.789*** | 0.639*** | - |
|  | (0.0913) | (0.0999) |  |
| **Age** | -0.000836 | 0.00506 | -0.00333 |
|  | (0.00898) | (0.00377) | (0.00868) |
| **Age Squared** | 0.000132 | 8.59e-05* | 0.000187* |
|  | (0.000104) | (4.43e-05) | (0.000104) |
| **Occupation (Base Cat.: Elementary Occupation)** | | | |
| Legislators, Senior Officials and Managers | 0.107** | 0.374*** | 0.445*** |
|  | (0.0512) | (0.0215) | (0.147) |
| Professionals | 0.0944** | 0.292*** | 0.389*** |
|  | (0.0460) | (0.0191) | (0.0404) |
| Associate Professionals | 0.122*** | 0.0943*** | 0.459*** |
|  | (0.0441) | (0.0180) | (0.0318) |
| Clerks | 0.128*** | 0.0794*** | 0.269*** |
|  | (0.0453) | (0.0188) | (0.0325) |
| Service Workers and Shop & Market Sales Workers | 0.0324 | 0.0751*** | 0.120*** |
|  | (0.0286) | (0.0136) | (0.0263) |
| Skilled Agricultural and Fishery Workers | -0.229* | 0.0410 | 0.0684 |
|  | (0.130) | (0.0446) | (0.0605) |
| Craft and Related Trades Workers | 0.146*** | 0.139*** | 0.154*** |

| Explanatory Variables | (1) | (2) | (3) |
|---|---|---|---|
| | Under | Adequate | Over |
| | (0.0337) | (0.0142) | (0.0249) |
| Plant and Machine Operators and Assemblers | 0.157*** | 0.186*** | 0.186*** |
| | (0.0349) | (0.0145) | (0.0258) |
| **Location of Workplace (Base Cat.: Rural)** | | | |
| Urban | 0.0657** | 0.0608*** | 0.0935*** |
| | (0.0293) | (0.0114) | (0.0237) |
| No Fixed Location | 0.180*** | 0.151*** | 0.141*** |
| | (0.0538) | (0.0249) | (0.0506) |
| **Contract (Base Cat.: Unwritten)** | | | |
| Written | 0.175*** | 0.189*** | 0.138*** |
| | (0.0218) | (0.00853) | (0.0173) |
| **Gender (Base Cat.: Male)** | | | |
| Female | -0.479*** | -0.392*** | -0.185*** |
| | (0.0536) | (0.0233) | (0.0557) |
| **Marital Status (Base Cat.: Unmarried)** | | | |
| Married | 0.130*** | 0.0924*** | 0.0220 |
| | (0.0330) | (0.0125) | (0.0250) |
| Others | 0.173*** | 0.0829*** | -0.0336 |
| | (0.0512) | (0.0252) | (0.0638) |
| **Social Group (Base Cat.: Scheduled Tribe)** | | | |
| Scheduled Caste | -0.0243 | -0.0614** | -0.0798 |
| | (0.0530) | (0.0246) | (0.0556) |
| OBC | -0.0250 | -0.0773*** | -0.0771** |
| | (0.0327) | (0.0150) | (0.0346) |
| Others | 0.0458 | 0.00359 | 0.0258 |
| | (0.0362) | (0.0166) | (0.0383) |
| **Religion (Base Cat.: Hindu)** | | | |
| Muslim | -0.0351 | -0.00608 | -0.00426 |
| | (0.0248) | (0.0114) | (0.0263) |
| Christian | -0.0125 | 0.0218 | -0.0180 |
| | (0.0412) | (0.0167) | (0.0369) |
| Others | 0.0331 | 0.00272 | 0.0989*** |
| | (0.0409) | (0.0170) | (0.0365) |
| **Sector (Base Cat.: Rural)** | | | |

| Explanatory Variables | (1) | (2) | (3) |
|---|---|---|---|
| | Under | Adequate | Over |
| Urban | 0.125** | 0.144*** | 0.145*** |
| | (0.0509) | (0.0226) | (0.0508) |
| **Selection term** | | | |
| Work | -0.0946** | -0.0183 | -0.108** |
| | (0.0457) | (0.0189) | (0.0444) |
| Wage/salaried Employment | -0.00607 | -0.0860 | -0.127 |
| | (0.162) | (0.0730) | (0.158) |
| Number of Observations | 5,599 | 27,301 | 5,542 |
| R-squared | 0.582 | 0.583 | 0.507 |

Source: Authors' calculation based on Periodic Labour Force Survey, 2018-19.
Note: (i) *** signals significant at 1% level, ** signals significant at 5% level, and * signals significant at 10% level.
 (ii) Robust standard errors are in parenthesis.
 (iii) The analysis also controls for broad industry groups, interaction of gender and marital status, and 35 states and union territories.

Table A7: Formal-informal sector wage gap: by match status

| Explanatory Variables | (1) Under | (2) Adequate | (3) Over |
|---|---|---|---|
| **Sector (Base Cat.: Formal)** | | | |
| Informal | -0.124*** | -0.187*** | -0.278*** |
|  | (0.0199) | (0.00921) | (0.0191) |
| **Education (Base Cat.: No Formal Schooling)** | | | |
| Primary | 0.116*** | 0.147 | - |
|  | (0.0278) | (0.0987) |  |
| Middle | 0.242*** | 0.263*** | -0.273*** |
|  | (0.0450) | (0.0987) | (0.0903) |
| Secondary | 0.465*** | 0.322*** | -0.250*** |
|  | (0.0444) | (0.0988) | (0.0647) |
| Higher Secondary | 0.645*** | 0.516*** | -0.152*** |
|  | (0.0450) | (0.0988) | (0.0480) |
| Graduate or Above | 1.052*** | 0.822*** | - |
|  | (0.0935) | (0.101) |  |
| **Age** | -0.00545 | 0.00672* | 0.00353 |
|  | (0.00924) | (0.00391) | (0.00892) |
| **Age Squared** | 0.000230** | 0.000109** | 0.000138 |
|  | (0.000107) | (4.61e-05) | (0.000107) |
| **Occupation (Base Cat.: Elementary Occupation)** | | | |
| Legislators, Senior Officials and Managers | 0.0601 | 0.429*** | 0.423*** |
|  | (0.0522) | (0.0224) | (0.150) |
| Professionals | 0.0704 | 0.311*** | 0.452*** |
|  | (0.0469) | (0.0198) | (0.0414) |
| Associate Professionals | 0.127*** | 0.110*** | 0.526*** |
|  | (0.0450) | (0.0187) | (0.0327) |
| Clerks | 0.134*** | 0.105*** | 0.332*** |
|  | (0.0463) | (0.0196) | (0.0335) |
| Service Workers and Shop & Market Sales Workers | 0.0140 | 0.0915*** | 0.151*** |
|  | (0.0296) | (0.0142) | (0.0271) |
| Skilled Agricultural and Fishery Workers | -0.320** | 0.0544 | 0.0572 |
|  | (0.127) | (0.0453) | (0.0610) |
| Craft and Related Trades Workers | 0.146*** | 0.137*** | 0.162*** |
|  | (0.0351) | (0.0149) | (0.0258) |

| Explanatory Variables | (1) | (2) | (3) |
|---|---|---|---|
| | Under | Adequate | Over |
| Plant and Machine Operators and Assemblers | 0.161*** | 0.190*** | 0.226*** |
| | (0.0361) | (0.0151) | (0.0266) |
| **Location of Workplace (Base Cat.: Rural)** | | | |
| Urban | 0.0851*** | 0.0748*** | 0.106*** |
| | (0.0301) | (0.0119) | (0.0247) |
| No Fixed Location | 0.161*** | 0.135*** | 0.140*** |
| | (0.0550) | (0.0254) | (0.0515) |
| **Contract (Base Cat.: Unwritten)** | | | |
| Written | 0.401*** | 0.390*** | 0.327*** |
| | (0.0206) | (0.00831) | (0.0164) |
| **Gender (Base Cat.: Male)** | | | |
| Female | -0.519*** | -0.456*** | -0.242*** |
| | (0.0551) | (0.0241) | (0.0568) |
| **Marital Status (Base Cat.: Unmarried)** | | | |
| Married | 0.149*** | 0.120*** | 0.0346 |
| | (0.0337) | (0.0129) | (0.0261) |
| Others | 0.190*** | 0.0778*** | -0.0647 |
| | (0.0521) | (0.0261) | (0.0658) |
| **Social Group (Base Cat.: Scheduled Tribe)** | | | |
| Scheduled Caste | -0.0583 | -0.0943*** | -0.120** |
| | (0.0533) | (0.0257) | (0.0579) |
| OBC | -0.0491 | -0.101*** | -0.0950*** |
| | (0.0336) | (0.0157) | (0.0357) |
| Others | 0.0217 | -0.000299 | 0.00274 |
| | (0.0369) | (0.0173) | (0.0397) |
| **Religion (Base Cat.: Hindu)** | | | |
| Muslim | -0.0288 | -0.0178 | -0.00316 |
| | (0.0254) | (0.0119) | (0.0271) |
| Christian | -0.0230 | 0.0371** | -0.0552 |
| | (0.0425) | (0.0174) | (0.0381) |
| Others | 0.0308 | -0.00730 | 0.140*** |
| | (0.0425) | (0.0178) | (0.0381) |
| **Sector (Base Cat.: Rural)** | | | |
| Urban | 0.125** | 0.134*** | 0.144*** |
| | (0.0506) | (0.0235) | (0.0532) |

| Explanatory Variables | (1) | (2) | (3) |
| --- | --- | --- | --- |
|  | Under | Adequate | Over |
| **Selection term** | | | |
| Work | -0.103** | -0.00501 | -0.0787* |
|  | (0.0471) | (0.0196) | (0.0450) |
| Wage/salaried Employment | 0.0326 | -0.00599 | -0.0655 |
|  | (0.159) | (0.0763) | (0.166) |
| Number of Observations | 6,032 | 29,055 | 5,870 |
| R-squared | 0.509 | 0.512 | 0.433 |

Source: Authors' calculation based on Periodic Labour Force Survey, 2018-19.
Note: (i) *** signals significant at 1% level, ** signals significant at 5% level, and * signals significant at 10% level.
 (ii) Robust standard errors are in parenthesis.
 (iii) The analysis also controls for broad industry groups, interaction of gender and marital status, and 35 states and union territories.